\documentclass[a4paper,11pt,showpacs]{revtex4}

\usepackage{amsmath} 
\usepackage{amsfonts}
\usepackage{graphicx}
\usepackage{dcolumn}
\usepackage{empheq}
\usepackage{upgreek}
\usepackage{latexsym,slashed}
\usepackage{amssymb,amsfonts}
\usepackage{geometry}
\usepackage{subfigure}
\usepackage[latin1]{inputenc}
\usepackage{color}
\usepackage{float}
\usepackage{makecell}
\usepackage{multirow,bigdelim}

\usepackage{cancel}
\usepackage{bm} %% bold mathematics
\usepackage{amsmath,mathrsfs} %% non gradito a latex2html!

\usepackage[parfill]{parskip}
\usepackage{booktabs} % for much better looking tables
\usepackage{array} % for better arrays (eg matrices) in maths
\usepackage{paralist} % very flexible & customisable lists (eg. enumerate/itemize, etc.)
\usepackage{verbatim} % adds environment for commenting out blocks of text & for better verbatim
\usepackage{subfig} % make it possible to include more than one captioned figure/table in a single float
\usepackage{multirow}
\usepackage[nottoc,notlof,notlot]{tocbibind} % Put the bibliography in the ToC
\usepackage[titles,subfigure]{tocloft} % Alter the style of the Table of Contents

\begin{document}
\tolerance=5000

\title{Dynamical analysis of a first order theory of bulk viscosity} 

\author{Giovanni Acquaviva}
\email{gioacqua@gmail.com}
\affiliation{Institute of Theoretical Physics, Faculty of Mathematics and Physics, Charles University in Prague, 18000 Prague, Czech Republic}
\author{Aroonkumar Beesham}
\email{abeesham@yahoo.com}
\affiliation{Department of Mathematical Sciences, University of Zululand, Private Bag X1001, Kwa-Dlangezwa 3886, South Africa}

\begin{abstract}
 We perform a global analysis of curved Friedmann-Robertson-Walker cosmologies in the  presence of a viscous fluid.  The fluid's bulk viscosity is governed by a first order theory recently proposed in \cite{disco2015}, and the analysis is carried out in a compactified parameter space with dimensionless coordinates.  We provide stability properties, cosmological interpretation and thermodynamic features of the critical points.
\end{abstract}

\pacs{98.80.-k, 95.36.+x}

\maketitle

\section{Introduction}

Most studies in cosmology are based on the assumption that the matter content of the Universe is well approximated by a perfect fluid description, {\it i.e.}, one without viscosity nor heat conduction. However there are stages in the evolution of the universe when viscosity and entropy-producing processes are expected to be important, especially during the early Universe: the reheating at the end of inflation, the decoupling of neutrinos from the primordial plasma, the nucleosynthesis and the decoupling of photons from matter during the recombination era.  At the same time, the analysis of data from the recent Planck survey \cite{ade2015} confirms a background geometry which, at very large scales, is isotropic and homogeneous (see also \cite{saadeh2016} for an independent analysis on the same dataset).  While shear viscosity and heat fluxes are related to the presence of anisotropies and inhomogeneities, bulk viscosity is related just to the kinematical expansion of the fluid's flow: the observational evidence hence justifies the choice of considering the cosmological effect of bulk viscosity alone among the possible dissipative processes, as long as the description is restricted to very large scales.\\

The role of bulk viscosity in cosmology has been explored from many points of view.  It has been proposed long ago as a way to avoid the initial Big Bang singularity \cite{Murphy} and to obtain a better understanding of the singularity itself \cite{Golda}.  As a phenomenological model of dissipative processes, bulk viscosity should describe the effect of isotropic expansion on the thermodynamic properties of fluids: for instance, one could interpret the dissipation as the result of a friction between different matter species undergoing a common Hubble expansion.  At the same time, bulk viscosity can provide a phenomenological description of particle creation in strong gravitational fields \cite{Barrow}.  One of the main features of viscosity is the possibility to lower the total effective pressure of the fluid to negative values: given that the Universe is currently undergoing accelerated expansion, and given that such effect in GR can be obtained by a sufficiently negative pressure of the matter source, bulk viscous fluids have been proposed as possible candidates for dark energy.  In this respect, it is worth mentioning that the traditional first-order model of bulk viscosity due to Eckart \cite{eckart} is formally equivalent to a generalized Chaplygin gas for some ranges of the free parameters involved.  Apart from cosmology, bulk viscosity could play a substantial role as well in astrophysical scenarios, such as in the growth of inhomogeneities that seed large-scale structures \cite{acqua2016,barbosa2017} and in the gravitational collapse of compact objects \cite{santos1984,herrera2009}.  However, in these cases the other dissipative processes are expected to contribute as well.\\

The formulation of a relativistic theory of dissipative processes dates back to the studies of Eckart \cite{eckart}, who introduced viscous contributions to the stress-energy tensor as functions of the four-velocity and the thermodynamical variables of the fluid.  In \cite{hiscock1985} Hiscock and Lindblom provided general arguments showing that a wide class of first-order theories are unstable: Eckart's approach, which considers first-order deviations from thermodynamic equilibrium, falls into such a class, and hence presents several shortcomings in terms of stability and physical viability.  The most evident of such problems is the superluminal propagation of signals.  In order to address the ensuing issue of causality, Mueller, Israel and Stewart \cite{mueller,israel} considered a formulation based on second-order deviations from equilibrium, in which the dissipative variables are treated as independent quantities satisfying their own evolution equations.  During the 1980s, the formulation of such {\it extended irreversible thermodynamics} has been deeply refined \cite{pavon1982,jou1988} and several applications in cosmology have been presented \cite{pavon1991,zimdahl1997}.  However, it has been argued \cite{disco2015} that the MIS theory is not uniquely defined, as the equations of motion that such variables satisfy rely on a certain degree of arbitrariness, and that its causal character is not yet fully understood.  Nonetheless, both Eckart's and MIS approaches (including possible modifications, {\it e.g.}, \cite{maart}) have been applied extensively to cosmology and their impact on the accelerated expansion of universe \cite{zimd2001,coli2007,li2009,gagnon2011,acqua2014,acqua2015} and on the formation of structures \cite{piatt2011,acqua2016} has been analysed. \\ 

We consider here a recent first-order fomulation of relativistic dissipative processes \cite{disco2017}, based on a previous approach due to Lichnerowicz \cite{lichne} and which does not introduce new, independent variables for the viscous quantities.  Although first-order in nature, such an approach does not fall into the class of theories that were proven unstable by Hiscock and Lindblom.  We present the basic structure of the theory in Sec.~\ref{equat}, specializing to a curved Friedmann-Robertson-Walker (FRW)  background metric and considering a single viscous fluid.  In Sec.~\ref{dynsyst} we recast the equations in the form of an autonomous dynamical system and analyse the general features of its critical elements from the cosmological point of view.  In Sec.~\ref{viscrad}, we further specialize to the case of viscous radiation and show the behaviour of the trajectories in the compactified parameter space.  Finally, in Sec.\ref{concl} we discuss and comment on the results.

\section{Equations}\label{equat}

The viscous energy-momentum tensor proposed in \cite{disco2015} has the form
\begin{equation}
 T_{\mu\nu} = \rho\, u_{\mu}u_{\nu} + \Big(p - \zeta\, \nabla_{\alpha}C^{\alpha} \Big)\, \left( g_{\mu\nu}+u_{\mu}u_{\nu} \right)\, ,\label{emt}
\end{equation}
where $\rho$ and $p$ are, respectively, the energy density and equilibrium pressure of the fluid.  The equilibrium pressure is modified by the bulk viscous term, where $\zeta\geq0$ is the bulk viscosity, $C^{\alpha}:=F\, u^{\alpha}$ is the dynamical velocity (or canonical momentum) of a fluid element and $F$ is the specific enthalpy of the fluid,
\begin{equation}
 F = \frac{\rho+p}{\mu}\, ,\label{F}
\end{equation}
with $\mu$ its rest mass density.  The adoption of the dissipative source in the form of eq.\eqref{emt} can be supported by the natural requirement that all the information about the properties of matter should be conveyed only by the structure of the energy-momentum tensor, without the need of introducing additional assumption on the dynamics of the variables involved -- which is instead the case in MIS formulation.  The dynamical velocity $C^\alpha$ appearing in the energy-momentum tensor has been considered as a suitable relativistic generalization of the concept of fluid's velocity in presence of dissipation (see \cite{disco2015} for a discussion).  It is worth noticing that $C^\alpha$ plays the role of {\it canonical momentum} in a Hamiltonian formulation of relativistic nonisentropic flows \cite{markakis2017}.  The bulk viscosity of the fluid is measured by $\nabla_{\alpha}C^{\alpha}$, that is the kinematical expansion of such vector field; the definition is compatible with the notion that bulk viscosity vanishes for an incompressible fluid, for which $\nabla_{\alpha}C^{\alpha}=0$.  A more general dissipative source would include also shear viscosity in terms of the spatial projection of the symmetrized quantity $\nabla_{(\alpha}C_{\beta)}$: however, in an isotropic background such contribution plays no role and for the purpose of the present analysis we will discard it.  In a FRW background we have
\begin{equation}\label{nablac}
 \nabla_{\alpha}C^{\alpha} = \dot{F} + 3\, H\, F\, ,
\end{equation}
where the dot is the derivative with respect to cosmic time.  Moreover, the rest mass is conserved along the flow lines $\nabla_{\alpha}\left( \mu u^{\alpha} \right)=0$, so that $\mu=\mu_0\, a^{-3}$.  The field and continuity equations obtained with the viscous energy-momentum tensor \eqref{emt} are the following:
\begin{align}
 & H^2 + \frac{k}{a^2} = \frac{1}{3}\, \rho\, ,\label{fried}\\
 &2\,\dot{H} +3\, H^2 + \frac{k}{a^2} = -\omega\, \rho + \zeta\, \nabla_{\alpha} C^{\alpha}\, , \label{ray}\\
 &\dot{\rho} + 3 H\, \left(1+\omega\right) \rho - 3\, H\, \zeta\, \nabla_{\alpha} C^{\alpha} = 0\, ,\label{cons}
\end{align}
where we have already implemented the barotropic equation of state $p=\omega\, \rho$ for the equilibrium pressure.  Moreover, it is usual to assume a generic power-law dependence of the bulk viscosity on the energy density of the fluid:
\begin{equation}
 \zeta = \zeta_0\, \rho^{\alpha}\quad \text{with}\quad \zeta_0\geq0\, .\label{zeta}
\end{equation}
Note that, in an expanding universe ($H>0$), positivity of entropy production is related to positivity of eq.\eqref{nablac}, because \cite{disco2015}
\begin{equation}
 T\, \nabla_{\beta}S^{\beta} = 3 \zeta\, H\, \nabla_{\alpha} C^{\alpha}\, ,
\end{equation}
where $T>0$ is the temperature of the fluid.  This will be relevant for determining the portion of the parameter space where entropy production is non-negative.

\section{Dynamical system}\label{dynsyst}

Dealing with a curved FRW background, we know that positive spatial curvature can lead to bouncing/recollapsing scenarios, whereas a negative or vanishing curvature is either always expanding or always collapsing.  The usual definition of {\it expansion-normalized variables} breaks down in the former case whenever $H=0$ at finite times.  For this reason we divide the analysis in two parts, the $k\leq0$ case and the $k>0$ case, as they require different definitions of dimensionless variables.  During the whole section, however, we keep a general EoS parameter $\omega \in (-1,1)$ and we will specialize to viscous radiation $\omega=1/3$ only in section \ref{viscrad}.

\subsection{Non-positive spatial curvature}\label{negcurv}

Imposing $k\leq0$ means that the term $k/a^2$ in the field equations is non-positive.  Hence we can safely define the new variables in the following way:
\begin{align}
 \Omega_{\rho} &= \frac{\rho}{3\, H^2}\, ,\label{omegarho}\\
 \Omega_k &= \frac{|k|}{3\, H^2\, a^2}\, ,\label{omegak}\\
 \Omega_C &= \frac{\zeta\, \nabla_{\alpha} C^{\alpha}}{\rho}\, .\label{omegac}
\end{align}
In terms of such variables, the Friedmann and Raychaudhuri equations respectively take the form
\begin{align}
 1 &= \Omega_{\rho} + \Omega_k\, , \label{fried2}\\
 \frac{\dot{H}}{H^2} &= \frac{3}{2}\, \Omega_{\rho}\, \Big[ \Omega_C - (1+\omega) \Big]\, . \label{ray2}
\end{align}
The first equation is a constraint that allows us to disregard the evolution of, {\it e.g.}, $\Omega_k$.  Such a constraint tells us that $\Omega_{\rho}\in [0,1]$.  The variable $\Omega_C$ instead is unbounded both from above and from below, which means that some trajectories of the system might escape to infinity: in order to capture such asymptotic behaviour, we define the new variable $X=\arctan \Omega_C$, such that $X\in[\pi/2,\pi/2]$.  We define as well the new evolution parameter $\tau=\log a(t)$. The derivative with respect to $\tau$ will be denoted by a prime and its relation with the cosmic time derivative is such that $X'=H^{-1}\dot{X}$.  Taking the prime derivative of the definitions of the relevant variables $\Omega_{\rho}$ and $X$, and making use of eqs.\eqref{cons} and \eqref{ray2}, we arrive at the following autonomous system:
\begin{align}
 \Omega_{\rho}' &= -3\, \Omega_{\rho}\, \left( 1-\Omega_{\rho} \right) \Big[ (1+\omega) - \tan X \Big]\, ,\\
 X' &=- \frac{3}{2(1-\omega)}\, \cos X\, \sin X\, \Big( 1-\omega +\tan X \Big) \Big[ (2\alpha+\Omega_{\rho})\, \left( 1+\omega-\tan X \right) - 2 \Big]\, .
\end{align}
The entropy production is related to the compact variables by
\begin{equation}\label{entprod_neg}
 T\, \nabla_{\beta}S^{\beta} = 9\, H^3\, \Omega_{\rho}\, \tan X\, .
\end{equation}
This means that for expanding models ($H>0$), the trajectories in the portion of the parameter space given by $X\in(0,\pi/2)$ correspond to dynamics with positive entropy production; conversely, in collapsing models ($H<0$), trajectories with positive entropy production have $X\in(-\pi/2,0)$.  The dividing trajectory $X=0$ is an invariant subset of the system, so we cannot expect positive entropy-producing initial conditions to evolve into negative entropy-producing states.

The critical points $P=\{X^*,\Omega_{\rho}^*\}$ of the system are easily found by solving the system
\begin{align}
 X' \left(X^*,\Omega_{\rho}^*\right) &= 0\\
 \Omega_{\rho}' \left(X^*,\Omega_{\rho}^*\right) &= 0\, ,
\end{align}
and they are given by
\begin{align*}
 P_0=\{0\, ,\, 0\}\quad &,\quad P_1=\{0\, ,\, 1\}\\
 P_2=\{-\arctan(1-\omega)\, ,\, 0\}\quad &,\quad P_3=\{-\arctan(1-\omega)\, ,\, 1\}\\
 P_4=\{\arctan(1+\omega-1/\alpha)\, ,\, 0\}\quad &,\quad P_5=\{\arctan(1+\omega-2/(1+2\alpha))\, ,\, 1\}\, ,
\end{align*}

\setlength{\extrarowheight}{11pt}
\begin{table}
\begin{ruledtabular}
  \begin{tabular}{c||cccccc}
    & $P_0$ & $P_1$ & $P_2$ & $P_3$ & $P_4$ & $P_5$ \\\hline
   $\alpha<-1/2$ & saddle & source & sink & saddle & saddle & sink \\
   $-1/2<\alpha<0$ & saddle & source & sink & saddle & saddle & source\\
   $0<\alpha<\frac{1}{2}$ & saddle & source & sink & source & saddle & saddle\\
   $\frac{1}{2}<\alpha<\frac{1-\omega}{2(1+\omega)}$ & saddle & source & saddle & source & sink & saddle\\
   $\frac{1-\omega}{2(1+\omega)}<\alpha<\frac{1}{1+\omega}$ & saddle & saddle & saddle & source & sink & source\\
   $\alpha>\frac{1}{1+\omega}$ & sink & saddle & saddle & source & saddle & source
  \end{tabular}
\caption{\label{curvneg_omeganeg} Stability of the critical points of the system with $k\leq0$ and $-1<\omega<0$.}
\end{ruledtabular}
\end{table}

\setlength{\extrarowheight}{11pt}
\begin{table}
\begin{ruledtabular}
  \begin{tabular}{c||cccccc}
    & $P_0$ & $P_1$ & $P_2$ & $P_3$ & $P_4$ & $P_5$ \\\hline
   $\alpha<-1/2$ & saddle & source & sink & saddle & saddle & sink \\
   $-1/2<\alpha<0$ & saddle & source & sink & saddle & saddle & source\\
   $0<\alpha<\frac{1-\omega}{2(1+\omega)}$ & saddle & source & sink & source & saddle & saddle\\
   $\frac{1-\omega}{2(1+\omega)}<\alpha<\frac{1}{2}$ & saddle & saddle & sink & source & saddle & source\\
   $\frac{1}{2}<\alpha<\frac{1}{1+\omega}$ & saddle & saddle & saddle & source & sink & source\\
   $\alpha>\frac{1}{1+\omega}$ & sink & saddle & saddle & source & saddle & source
  \end{tabular}
\caption{\label{curvneg_omegapos} Stability of the critical points of the system with $k\leq0$ and $0<\omega<1$.}
\end{ruledtabular}
\end{table}

For purposes of clarity, their stability properties for the case of expanding models ($H>0$) are listed in Table \ref{curvneg_omeganeg} for $-1<\omega<0$ and in Table \ref{curvneg_omegapos} for $0<\omega<1$.  The case of collapsing models can be obtained by swapping the roles of sinks and sources.  The {\it deceleration parameter} $q$ and the {\it effective EoS parameter} $\omega_E$ are given by  
\begin{align}
 q &\equiv -1-\frac{\dot{H}}{H^2} = \frac{3}{2}\, \Omega_{\rho}\, \Big[ (1+\omega) - \tan X \Big] - 1\, ,\\
 \omega_E &\equiv \frac{p - \zeta\, \nabla_{\alpha}C^{\alpha}}{\rho} = \omega - \tan X\, ,
\end{align}
while the scale factor evolution can be obtained by integrating eq.\eqref{ray2} in the critical points.

Points $P_0$, $P_2$ and $P_4$ are vacuum models with exponential expansion of the scale factor ($q=-1$) and they can act either as transients or as future attractors in expanding situations.  The point $P_1$ is an inviscid fluid solution with $\omega_E=\omega$ and $q=(1+3\omega)/2$ and it can be either a transient saddle or a past attractor.  The point $P_3$ is a stiff matter model with $q=2$ and $\omega_E=1$, with possible character of future attractor.  Finally, point $P_5$ has an effective equation of state which depends explicitly on the viscosity parameter as $\omega_E = \frac{1-2\alpha}{1+2\alpha}$; it can represent a phantom model when $\alpha<-1/2$, in which case is a future attractor for the system.

\subsection{Positive spatial curvature}\label{poscurv}

If $k>0$ then one is faced with the possibility of bouncing or recollapsing models, in which cases the dimensionless variables defined in the previous section are ill-defined in the turning points of the scale factor, {\it i.e.} when $H=0$.  However the quantity
\begin{equation}
 D = \sqrt{H^2+\frac{k}{a^2}}
\end{equation}
is always positive definite, so we can define the following normalized variables:
\begin{align}
 \Omega_H &= \frac{H}{D}\, ,\\
 \Omega_{\rho} &= \frac{\rho}{3\, D^2}\, ,\\
 \Omega_C & = \frac{\zeta\, \nabla_{\alpha}C^{\alpha}}{3\, H\, D}\, .
\end{align}
Again we define the compactified variable $X=\arctan \Omega_C$, so that the Friedmann constraint and Raychaudhuri equation read
\begin{align}
 \Omega_{\rho} &= 1\, ,\\
 \frac{\dot{H}}{H^2} &= -\frac{1}{2\, \Omega_H^2}\, \left[ 1 +2\, \Omega_H^2 + 3\, (\omega-\Omega_H \tan X) \right]\, .
\end{align}
We see immediately that $\Omega_{\rho}$ is not a dynamical degree of freedom.  The variable $\Omega_H$ is defined in the interval $[-1,1]$ and its sign is positive/negative iff the metric is expanding/contracting; the boundary values $\Omega_H=\pm1$ represent the expanding/contracting spatially flat cases.  An additional useful equation is given by the evolution of $D$:
\begin{equation}
 \frac{\dot{D}}{D^2} = -\frac{3}{2}\, \Omega_H\, \Big[ 1+\omega-\Omega_H \tan X \Big]\, .
\end{equation}
Analogously to the previous case, we define the new time derivative $X'=D^{-1}\dot{X}$, so that the prime derivatives of the dynamical variables give us the following system:
\begin{align}
 \Omega_H' &= -\frac{1}{2}\, \left(1-\Omega_H^2\right)\, \Big[ 1+3\, (\omega-\Omega_H \tan X) \Big]\, ,\\
 X' &= -\frac{\sin X}{2\, (1-\omega)}\, \Big[ 3\, (1-\omega) \left( \omega+2\alpha(1+\omega)-1 \right) \Omega_H \cos X\, + \Big.\\ 
 &\quad\, \Big. + \sin X \Big( 1+3\omega+\Omega_H^2 \left( 3\, (1+4\alpha)\, \omega-7 \right)-3\,\Omega_H\, \tan X \left( 1+2\, \alpha\, \Omega_H^2 \right) \Big) \Big]\, .
\end{align}

In this case, the entropy production is given by
\begin{equation}\label{entprod_pos}
T\, \nabla_{\alpha}S^{\alpha} = 9\, H^3\, \frac{\tan X}{\Omega_H}\, .
\end{equation}
Hence, in the parameter space spanned by $\left( X, \Omega_H \right)$, entropy production is positive iff $X\in(0,\pi/2)$.  Also in this case the subspace $X=0$ is an invariant subset of the system, so trajectories starting with positive entropy production cannot cross to the negative entropy production part.

\setlength{\extrarowheight}{13pt}
\begin{table}
\begin{ruledtabular}
  \begin{tabular}{c||cccccc}
    & $Q_0^+$ & $Q_0^-$ & $Q_1^+$ & $Q_1^-$ & $Q_2^+$ & $Q_2^-$ \\\hline
   $\alpha<-1/2$ & saddle & saddle & saddle & saddle & sink & source \\
   $-1/2<\alpha<0$ & saddle & saddle & saddle & saddle & source & sink\\
   $0<\alpha<1$ & saddle & saddle & source & sink & saddle & saddle\\
   $1<\alpha<\frac{1}{2}\frac{1-\omega}{1+\omega}$ & saddle & saddle & source & sink & sink & source\\
   $\alpha>\frac{1}{2}\frac{1-\omega}{1+\omega}$ & $\left\{\parbox{2.5cm}{sink $X\rightarrow0^-$\\ saddle $X\rightarrow0^+$}\right.$ & $\left\{\parbox{2.5cm}{saddle $X\rightarrow0^-$\\ source $X\rightarrow0^+$}\right.$ & source & sink & saddle & saddle
  \end{tabular}
\caption{\label{poscrit1} Stability of the finite critical points of the system with $k>0$ and $-1<\omega<-1/3$.}
\end{ruledtabular}
\end{table}

\setlength{\extrarowheight}{11pt}
\begin{table}
\begin{ruledtabular}
  \begin{tabular}{c||cccccc}
    & $Q_0^+$ & $Q_0^-$ & $Q_1^+$ & $Q_1^-$ & $Q_2^+$ & $Q_2^-$ \\\hline
   $\alpha<-1/2$ & source & sink & saddle & saddle & sink & source \\
   $-1/2<\alpha<0$ & source & sink & saddle & saddle & source & sink\\
   $0<\alpha<\frac{1}{2}\frac{1-\omega}{1+\omega}$ & source & sink & source & sink & saddle & saddle\\
   $\frac{1}{2}\frac{1-\omega}{1+\omega}<\alpha<1$ & saddle & saddle & source & sink & source & sink\\
   $\alpha>1$ & saddle & saddle & source & sink & saddle & saddle
  \end{tabular}
\caption{\label{poscrit2} Stability of the finite critical points of the system with $k>0$ and $-1/3<\omega<1$.}
\end{ruledtabular}
\end{table}

The critical points are calculated as before and they are given by
\begin{align*}
 Q_0^+=\{0\, ,\, 1\}\quad &,\quad Q_0^-=\{0\, ,\, -1\}\\
 Q_1^+=\{-\arctan(1-\omega)\, ,\, 1\}\quad &,\quad Q_1^-=\{-\arctan(1-\omega)\, ,\, -1\}\\
 Q_2^+=\{\arctan(1+\omega-1/\alpha)\, ,\, 1\}\quad &,\quad Q_2^-=\{\arctan(1+\omega-2/(1+2\alpha))\, ,\, -1\}\, ,
\end{align*}
Again, in order to present the results in a clear way, we separate the cases when $-1<\omega<-1/3$  and when $-1/3<\omega<1$, which are shown respectively in Tables \ref{poscrit1} and \ref{poscrit2}.  The cosmological parameters that characterize the critical points in this case are
\begin{align}
 q &= \frac{1}{2}\, \frac{1+3\, (\omega-\Omega_H\, \tan X)}{\Omega_H^2}\, ,\\
 \omega_E &= \omega-\Omega_H\, \tan X\, .
\end{align}
The critical points $Q_0^{\pm}$ represent inviscid fluid models with $\omega_E=\omega$.  Points $Q_1^{\pm}$ are stiff matter solutions with $\omega_E=1$.  Points $Q_2^{\pm}$ have $\omega_E=\frac{1-2\alpha}{1+2\alpha}$ and $q=\frac{2(1-\alpha)}{1+2\alpha}$, so that they can represent phantom models for $\alpha<-1/2$.

\section{Viscous radiation}\label{viscrad}

\begin{figure*}[h!!!]
 \begin{center}
  \includegraphics[width=14cm]{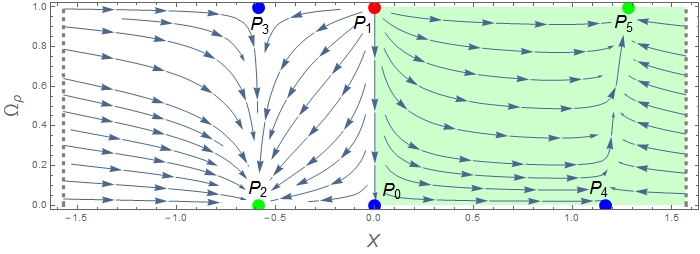}\vspace{1cm}
  \includegraphics[width=14cm]{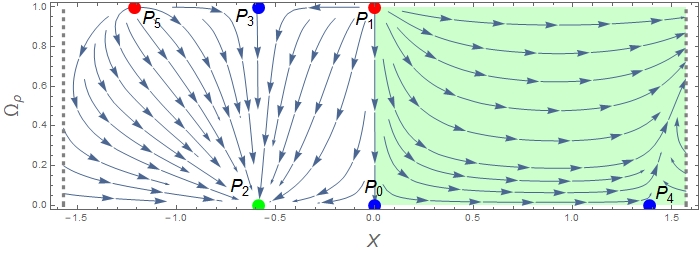}\vspace{1cm}
  \includegraphics[width=14cm]{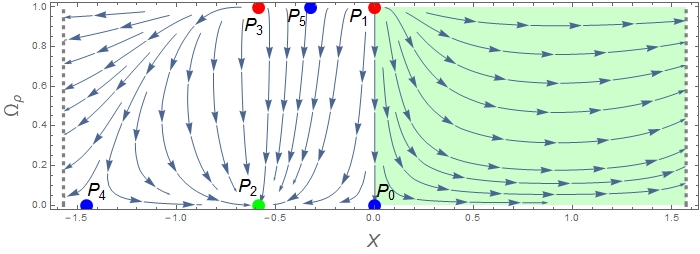}
 \caption{\label{neg_rad_1}  Trajectories in the parameter space for the system with $k\leq0$, $\omega=1/3$ and expanding dynamics ($H>0$).  From the top: $\alpha=-1$, $\alpha=-1/4$, $\alpha=1/10$.  Dots identify sources (red), saddles (blue) and sinks (green).  The green shaded area is the positive entropy production region.  The dashed line is the compactified boundary of the system.}
\end{center}
\end{figure*}

\begin{figure*}[h!!!]
 \begin{center}
  \includegraphics[width=14cm]{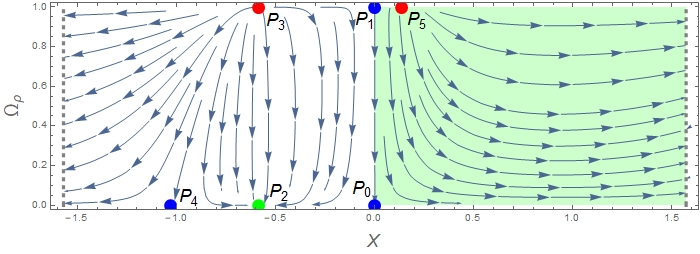}\vspace{1cm}
  \includegraphics[width=14cm]{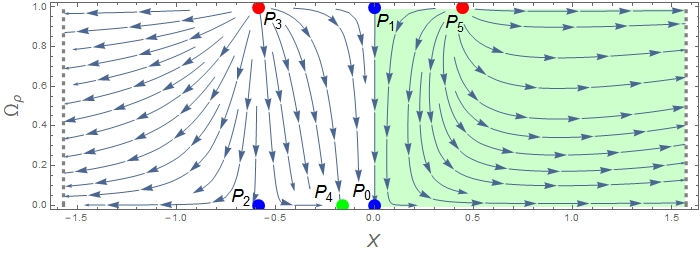}\vspace{1cm}
  \includegraphics[width=14cm]{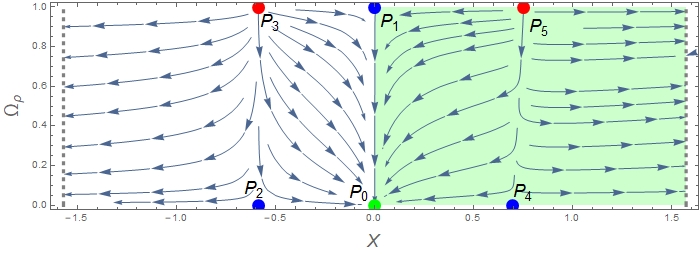}
 \caption{\label{neg_rad_2}  Trajectories in the parameter space for the system with $k\leq0$, $\omega=1/3$ and expanding dynamics ($H>0$).  From the top: $\alpha=1/3$, $\alpha=2/3$, $\alpha=2$.  Dots identify sources (red), saddles (blue) and sinks (green).  The green shaded area is the positive entropy production region.  The dashed line is the compactified boundary of the system.}
\end{center}
\end{figure*}

\begin{figure*}[h!!!]
 \begin{center}
  \includegraphics[width=8cm]{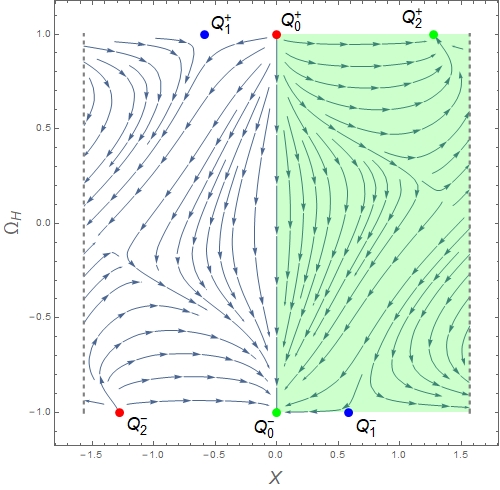}\hspace{0.3cm}
  \includegraphics[width=8cm]{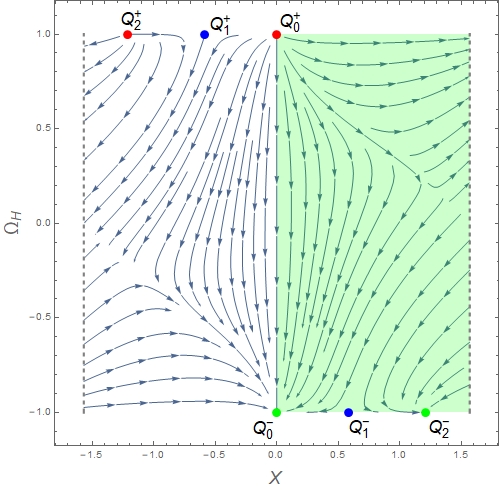}\vspace{1.7cm}
  \includegraphics[width=8cm]{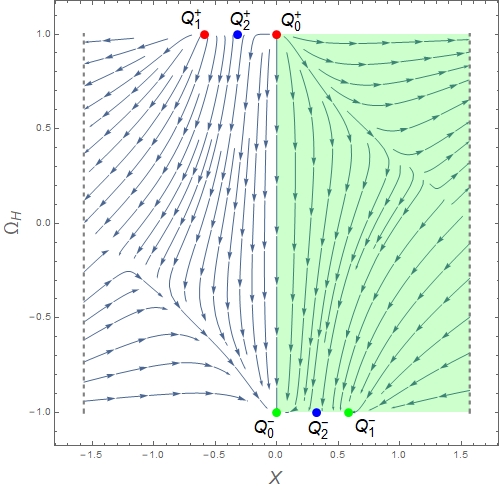}
 \caption{\label{pos_rad_1} Trajectories in the parameter space for the system with $k>0$ and $\omega=1/3$.  From top left: $\alpha=-1$, $\alpha=-1/4$, $\alpha=1/10$.  Dots identify sources (red), saddles (blue) and sinks (green).  The green shaded area is the positive entropy production region.  The dashed line is the compactified boundary of the system.}
\end{center}
\end{figure*}

\begin{figure*}[h!!!]
 \begin{center}
  \includegraphics[width=8cm]{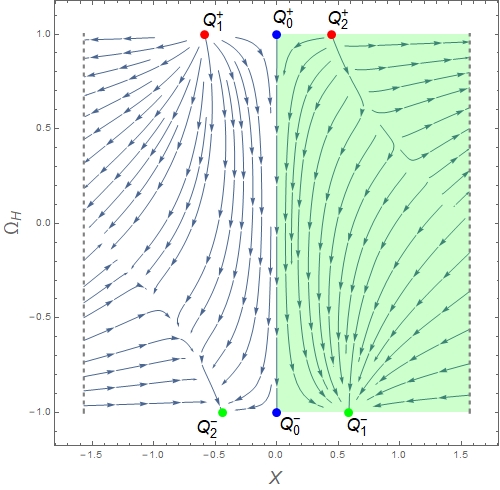}\hspace{0.3cm}
  \includegraphics[width=8cm]{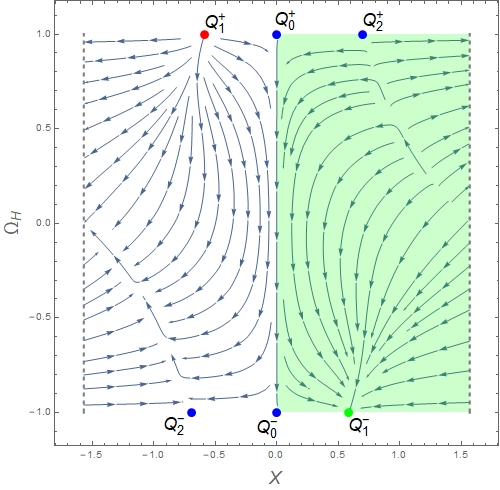}
 \caption{\label{pos_rad_2} Trajectories in the parameter space for the system with $k>0$ and $\omega=1/3$.  From left: $\alpha=2/3$ and $\alpha=3/2$.  Dots identify sources (red), saddles (blue) and sinks (green).  The green shaded area is the positive entropy production region.  The dashed line is the compactified boundary of the system.}
\end{center}
\end{figure*}

In order to visualize a physically meaningful case, we now specialize the analysis to the case of viscous radiation $(\omega=1/3)$.  We plot the trajectories in the compactified parameter space in Fig.\ref{neg_rad_1} and Fig.\ref{neg_rad_2} for $k\leq0$, and in Fig.\ref{pos_rad_1} and Fig.\ref{pos_rad_2} for $k>0$.  We choose representative values of the parameter $\alpha$ corresponding to the ranges specified in the Tables of stability given in the previous section.  The dots in the plots identify sinks (green), saddles (blue) and sources (red).  The green shaded region in the plots corresponds to positivity of entropy production, according to eqs. \eqref{entprod_neg} and \eqref{entprod_pos}.  

We stress that in the case $k\leq0$ we have plotted only the expanding ($H>0$) portion of the system: the collapsing part can be obtained by inverting the direction of the flows; moreover, in this case the positive entropy production regions will be $X<0$.  As noted before, $X=0$ is an invariant subset, so the sign of entropy production is preserved during the dynamics.  The same happens in the $k>0$ case: during the dynamics the sign of entropy production is also preserved.


The behaviour of the system at infinity is now encoded in the boundaries $X=\pm\pi/2$, identified in the figures by a dashed line.  In both curvature cases, the trajectories reach such boundaries with a specific constant values of the bounded coordinate ($\Omega_{\rho}$ for $k\leq0$ and $\Omega_H$ for $k>0$), the value of which depends on the initial conditions of the dynamics.  In the non-positive curvature case, focusing on the expanding sector (Fig.\ref{neg_rad_1} and \ref{neg_rad_2}), an asymptotic constant value $\Omega_{\rho}^0\in(0,1)$ on the boundary allows us to calculate the scale factor by integrating the Friedmann constraint:
\begin{align}
 &a(t) = \sqrt{\frac{|k|}{3\left( 1-\Omega_{\rho}^0 \right)}}\, \left( t - t_{BB} \right)\, .
\end{align}
%\clearpage
This corresponds to an asymptotically Milne-like model, with Big Bang time $t_{BB} = -a_0\, \sqrt{3\left( 1-\Omega_{\rho}^0 \right)/|k|}$ and $a_0=a\left(t=0\right)$.  Consequently, the Hubble expansion and the energy density scale like
\begin{align}
 H &= \sqrt{\frac{|k|}{3\left(1-\Omega_{\rho}^0\right)}}\ a^{-1}\, ,\\
 \rho &= \frac{|k|\, \Omega_{\rho}^0}{1-\Omega_{\rho}^0}\ a^{-2}\,  .
\end{align}
By considering a usual functional dependence of the temperature of the fluid on its energy density, {\it i.e.} $T=T_0\, \rho^{\omega/(1+\omega)}$ with $\omega=1/3$, eventually one can evaluate the scaling of the entropy production in such asymptotic states:
\begin{equation}
 \nabla_aS^a = c_{\pm}\, \cdot\, a^{-2\left( \alpha+1/4 \right)}\, ,
\end{equation}
where $c_+$ (resp. $c_-$) is a positive (resp. negative) constant on the boundary $X_C=\pi/2$ (resp. $X_C=-\pi/2$).  For instance, in the positive entropy production part of the dynamics one has
\begin{itemize}
 \item if $\alpha+1/4>0$:
 \begin{itemize} 
  \item $\nabla_aS^a\rightarrow0$ for $a\rightarrow\infty$ (stable boundary)
  \item $\nabla_aS^a\rightarrow\infty$ for $a\rightarrow0$ (unstable boundary)
 \end{itemize}
 
 \item if $\alpha+1/4<0$:
 \begin{itemize} 
  \item $\nabla_aS^a\rightarrow\infty$  for $a\rightarrow\infty$ (stable boundary)
  \item $\nabla_aS^a\rightarrow0$ for $a\rightarrow0$ (unstable boundary)
 \end{itemize}
\end{itemize}
An analogous analysis applies in the case $k>0$, where a constant value $\Omega_H^0$ on the compactified boundaries allows to calculate the scale factor by integrating the definition of the variable itself:
\[
    a(t) = \left\{\begin{array}{lr}
        \sqrt{\frac{k\, (\Omega_H^0)^2}{1-(\Omega_H^0)^2}}\, \left( t-t_{BB} \right), & \text{for } \Omega_H^0>0\\
        \sqrt{\frac{k\, (\Omega_H^0)^2}{1-(\Omega_H^0)^2}}\, \left( t_{BC}-t \right), & \text{for } \Omega_H^0<0\, ,
        \end{array}\right.
  \]
where $t_{BB}$ and $t_{BC}$ are the Big Bang and the Big Crunch times respectively.  Having the same asymptotic time-dependence of the scale factor as in the previous case, also the scaling properties of the entropy production are analogous.

\section{Conclusions}\label{concl}

In the present work we have analysed the system of Einstein's equations sourced by a single dissipative fluid in the context of a first-order theory of relativistic dissipation.  The generically curved FRW metric has been taken as cosmological background. The system has been recast in the form of a dimensionless, autonomous system of equations whose equilibrium points represent different dynamics of the scale factor.  The study of the stability of such critical points allowed us to assess the past and future behaviour of the system, characterized unambiguously by the deceleration and effective equation of state parameters.  The results obtained here are in accord with those presented in \cite{disco2017}, for instance regarding the attractor behaviour of the phantom solution for $\alpha<-1/2$ (critical point $P_5$), but as well highlight additional features: de Sitter-like future attractors exist for different ranges of the parameter $\alpha$ in the non-positive curvature case (points $P_0$, $P_2$ and $P_4$); for $\alpha>0$, a stiff matter-dominated solution is found as a past attractor for $k\leq0$ (point $P_3$) and as both a past and a future attractor for $k>0$ (points $Q_1^{\pm}$); we notice further that in general the evolution preserves the sign of the entropy production, so that positive entropy-producing initial conditions cannot evolve into negative entropy-producing states.  

It seems natural to ask whether it is possible to obtain an evolution that could interpolate between an inflationary epoch due to viscosity up to the later radiation-dominated phase: the results presented here indicate that it is not possible to have such a behaviour.  Indeed, if we require positive and non-divergent entropy production throughout the evolution, there is no trajectory connecting de Sitter-like solutions to radiation-dominated ones in an expanding dynamics.  Instead, under the same physical assumptions on entropic evolution, this model of viscous radiation seems to give rise quite easily to the opposite transition, from a radiation-dominated early epoch towards a late-time accelerated phantom behaviour.  In the non-positive curvature scenario, this is obtained for $\alpha<-1/2$ ({\it e.g.} trajectories connecting $P_1$ and $P_5$ in Fig.\ref{neg_rad_1}) and $\alpha>3/4$ ({\it e.g.} trajectories connecting $P_5$ and $P_0$ in Fig.\ref{neg_rad_2}), and the rate of acceleration is higher the closer $\alpha$ is to such bounds.  Considering positive curvature instead and requiring only non-recollapsing scenarios, the possibility to have such transition needs $\alpha<-1/2$.  These considerations suggest that the present model of viscosity (at least in the single-fluid scenario) is not suitable for describing an effective inflationary regime together with its exit mechanism towards reheating and subsequent radiation dominance.  Regarding instead the possibility of describing a late-time acceleration of the Universe -- which arises more naturally in the global dynamics -- the attractors discussed above have generically $\omega_E<-1$, with the bound reached only for $\alpha\rightarrow\pm\infty$.  On one hand, assuming a sufficiently big value for $|\alpha|$, one could think of including in the analysis an additional inviscid dust component which would likely dominate right after the radiation: the expected dynamics could then interpolate between radiation, matter and accelerated expansion phases.  On the other hand, a very big value of $|\alpha|$ that would ensure a late-time exponential expansion in accord with observations might instead have serious repercussions at perturbative level.  In this regard, an analysis of the growth of perturbations in this framework of dissipative processes would certainly be useful in assessing the viability of such scenario.

\begin{acknowledgments}
 GA acknowledges financial support from the Grant No.~17-16260Y of the Czech Science Foundation (GA\v{C}R).  The authors thank the referees for their constructive comments, which lead to significant improvements in the manuscript.
\end{acknowledgments}

\end{document}